# Quantitative X-ray Schlieren Nanotomography for Hyperspectral Phase and Absorption Imaging

Authors: Herve Hugonnet[1,2,†], KyeoReh Lee[1,2,4,†,*], Sugeun Jo[3], Jun Lim[3,*], and YongKeun Park[1,2,*]

[1]*Department of Physics, Korea Advanced Institute of Science and Technology, Daejeon, 34141, Republic of Korea.*
[2]*KAIST Institute for Health Science and Technology, Korea Advanced Institute of Science and Technology, Daejeon, 34141, Republic of Korea.*
[3]*Pohang Accelerator Laboratory, Pohang University of Science and Technology, Pohang, 37637, Republic of Korea.*
[4]*Current address: Department of Applied Physics, Yale University, New Haven, 06520, Connecticut, USA.*

[†] *These authors contributed equally to this work*
[*] *Corresponding author(s). E-mail(s): lee.kyeo@gmail.com; limjun@postech.ac.kr; yk.park@kaist.ac.kr*

**Abstract**

Hyperspectral X-rays imaging holds promise for three-dimensional (3D) chemical analysis but remains limited in simultaneously capturing phase and absorption information due to complex setups and data burdens. We introduce quantitative X-ray schlieren nanotomography (XSN), a simple, fast, and high-resolution X-ray phase imaging technique that overcomes these limitations. XSN employs a partially coherent illumination and a pupil-plane cutoff filter to encode directional phase contrast, enabling single-shot acquisition. A quasi-Newton iterative algorithm reconstructs quantitative phase and absorption images from intensity data, even under strong scattering conditions. Scanning across X-ray energies further allows four-dimensional imaging (3D spatial + spectral). We validate the method's accuracy and resolution on reference samples and apply it to lithium battery cathodes, visualizing nanoscale microcracks and mapping chemical compositions. XSN provides a robust framework for hyperspectral phase nanotomography with broad applicability across materials science, biology, and energy research.

**Introduction**

Hard X-ray imaging is widely used in industrial and scientific applications due to its strong penetrability and elemental specificity[1-3]. However, the weak interaction of X-rays with matter that underlies this penetrability also leads to low intrinsic contrast for fine features, especially at the microscopic scale. This lack of contrast has motivated the development of X-ray phase imaging techniques, which can provide higher signal-to-noise ratio for a given radiation dose and thereby reduce sample damage while increasing imaging speed[4]. These advantages have spurred extensive research into diverse X-ray phase imaging methods[1,5-9].

One widely used approach to X-ray phase contrast is Zernike phase contrast, as implemented in full-field transmission X-ray microscopy (TXM) and nanotomography[10-12]. Zernike phase contrast yields high image contrast but has notable drawbacks: the phase signal is rendered qualitatively (non-quantitatively), and the method requires specialized optics such as a grating condenser[13], a glass capillary, and a phase ring at the back focal plane of the objective. Alternative full-field phase imaging methods include grating interferometry and propagation-based phase contrast, but these typically demand multiple exposures or complex optical configurations to retrieve quantitative phase information[1,5-9]. Beyond full-field approaches, scanning coherent diffraction techniques such as ptychography can provide quantitative phase images with accuracy and nanoscale resolution[14]. Ptychographic phase tomography, however, requires collecting a large number of far-field diffraction patterns for each view, followed by intensive iterative reconstruction. This results in prohibitively long acquisition times and data processing loads for three-dimensional and spectral (multi-energy) analyses[15]. Consequently, there is a need for phase tomography methods that combine simplicity, speed, high resolution, and spectral flexibility.

To address this need, we previously introduced Kramers–Kronig (KK) nanotomography[16] as a way to achieve quantitative X-ray phase imaging with minimal modifications to a standard TXM. By inserting a cutoff filter at the pupil plane of a standard off-axis TXM, KK nanotomography enabled single-shot retrieval of phase from intensity images via the Kramers–Kronig relations. While effective in principle, the original KK nanotomography relied on spatially coherent illumination and assumed weak scattering, which imposed several practical limitations. In addition, the need for high spatial coherence led to strong speckle noise and a restricted field of view, and the additional spatial filtering of the beam reduced the photon flux available for imaging.

Here, we introduce quantitative X-ray schlieren nanotomography (XSN) to overcome the limitations of KK nanotomography and extend phase tomography to more general and practical imaging conditions. XSN retains the basic optical layout of KK nanotomography[16] but incorporates two key innovations inspired by schlieren imaging and advanced phase retrieval: a schlieren imaging-like incoherent illumination and asymmetric filtering scheme and a quasi-Newton iterative inversion algorithm. In XSN, we integrate these concepts into a full 3D tomographic framework: starting from an initial estimate, the algorithm iteratively updates the sample's complex refractive index distribution (phase and absorption) using a quasi-Newton optimization for fast convergence speed. This allows accurate imaging of strongly scattering or highly absorptive samples that would defy conventional single-step phase retrieval. The combination of schlieren-inspired phase contrast with a robust iterative solver yields quantitative reconstructions without requiring multiple frame recordings or prior weak-phase assumptions.

In this work, we validate the accuracy and spatial resolution of XSN using nanoscale test objects and demonstrate its utility on battery electrode materials. We show that XSN can visualize fine microcracks inside a lithium nickel–manganese–cobalt oxide (NMC) cathode particle and, by performing tomography at multiple X-ray energies, map the particle's three-dimensional chemical composition. These results showcase XSN's unique capability to integrate schlieren-inspired contrast, partially coherent illumination, and advanced iterative reconstruction—thereby enabling comprehensive nanoscale imaging of both structure and composition.

**Results**

**Quantitative X-ray schlieren nanotomography (XSN)**

XSN extends the conventional off-axis nanotomography setup by incorporating a pupil-plane cutoff filter and adopting a schlieren-inspired imaging scheme (Fig. 1a). This configuration allows phase gradients within the sample to be translated into detectable intensity variations in a single shot. A rotating diffuser is introduced to generate a partially coherent illumination, which effectively homogenizes the beam, reducing both speckle noise and out-of-focus fringe artifacts that are typical of fully coherent X-ray source[17,18].

To recover quantitative phase and absorption information, we developed an iterative reconstruction algorithm that explicitly models partial coherence in the illumination. This approach enables accurate retrieval of both phase and absorption distributions, even in samples exhibiting strong absorption or scattering (Fig. 1b).

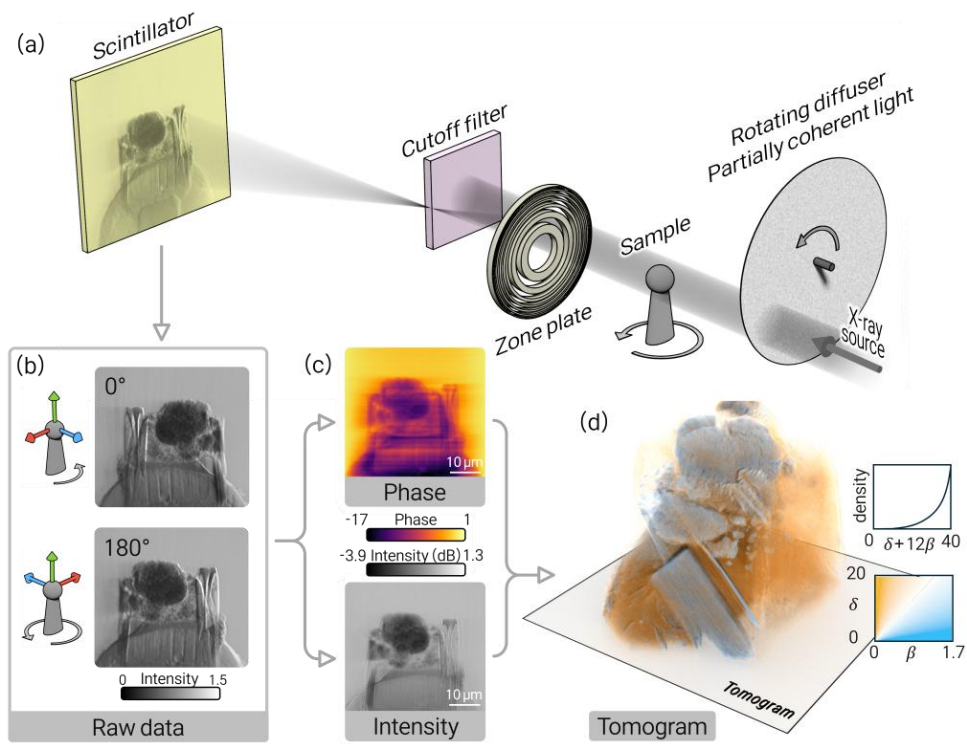

Figure 1. Graphical summary of the XSN method. (a) Schematic of the optical setup. A rotating diffuser generates partially coherent illumination, while a zone plate provides image magnification. A fixed pupil-plane cutoff filter introduces directional phase contrast. (b) Representative field reconstructions of a battery cathode particle at 0° and 180° projection angles. (c) Retrieved quantitative phase images. (d)

Reconstructed tomogram obtained via filtered back projection. Refractive index values ($\delta$ and $\beta$) are reported in units of $10^{-6}$.

The complex transmittance of a thin object can be expressed as $\exp[i\varphi(\mathbf{r}) + \alpha(\mathbf{r})]$, where $\varphi(\mathbf{r})$ and $\alpha(\mathbf{r})$ are phase and absorption coefficient, and $\mathbf{r}$ denotes the spatial coordinate. Under partially coherent illumination characterized by an angular power spectral density (PSD) $I_\mathbf{k}$ for a given illumination wave vector $\mathbf{k}$. The total transmitted intensity becomes

$$I(\mathbf{r}) = \int \left| \mathcal{F}^{-1}\left\{ P(\mathbf{u})\mathcal{F}\left\{ \sqrt{I_\mathbf{k}} e^{i\mathbf{k}\mathbf{r}} e^{\alpha(\mathbf{r})+i\varphi(\mathbf{r})} \right\} \right\} \right|^2 d\mathbf{k}, \tag{1}$$

where $\mathcal{F}\{\cdot\}$ is the Fourier transform and $P(\mathbf{u})$ is the pupil cutoff filtering function, and $\mathbf{u}$ is the spatial frequency vector corresponds to $\mathbf{r}$. This equation can be linearized (see Method and Ref.[19]) and reformulated as,

$$I = I_{tot} + \text{PSF}_\alpha * \alpha + \text{PSF}_\varphi * \varphi, \tag{2}$$

where $*$ indicates convolution, $I_{tot} = \int I_\mathbf{k} d\mathbf{k}$ is the background intensity in the absence of sample, $\text{PSF}_\varphi$ and $\text{PSF}_\alpha$ are the point spread functions (PSFs) for phase and absorption, respectively. These PSFs can be expressed as:

$$\begin{aligned} \text{PSF}_\alpha &= 4\,\text{Re}\left[ \mathcal{F}^{-1}\{I_\mathbf{k} P\}^* \mathcal{F}^{-1}\{P\} \right] \\ \text{PSF}_\varphi &= 4\,\text{Im}\left[ \mathcal{F}^{-1}\{I_\mathbf{k} P\}^* \mathcal{F}^{-1}\{P\} \right] \end{aligned}, \tag{3}$$

and the argument $\mathbf{r}$ is omitted in the equations for brevity.

According to Eq. (2), the intensity measured at the detector can be represented as a convolution of the object's phase and absorption distributions with their respective PSFs, $\text{PSF}_\varphi$ and $\text{PSF}_\alpha$. These PSFs depends on both the angular PSD $I_\mathbf{k}$, which characterizes the partial coherence of illumination, and the geometry of the pupil-plane filter $P$. Representative plots of $I_\mathbf{k}$, $P$, and the resulting $\text{PSF}_{\alpha,\varphi}$ under our experimental conditions are shown in Fig. S1a.

To independently recover the phase $\varphi(\mathbf{r})$ and absorption $\alpha(\mathbf{r})$ components from Eq. (2), at least two independent measurements with different illumination conditions or pupil filters are typically required. In XSN, this requirement is met without altering the optical setup: by rotating the sample by 180°, the mirrored configuration effectively generates a complementary measurement, with the direction of the phase gradient reversed relative to the fixed cutoff filter (Fig. 1b). This mirrored intensity profile provides the additional constraint necessary to decouple phase and absorption signals without any physical modification of the optical system (see Fig. 1c and Eq. (4)).

By utilizing two mirrored experimental intensity measurements, $I_\text{exp}^0$ and $I_\text{exp}^\pi$, acquired with sample rotations differing by 180°, Eq. (2) can be inverted through deconvolution with Tikhonov regularization[20]:

$$\begin{bmatrix} \alpha \\ \varphi \end{bmatrix} \simeq \mathcal{F}^{-1}\left\{ \mathbf{D}\mathcal{F}\left\{ \begin{pmatrix} I_\text{exp}^0 - I_\text{tot}^0 \\ I_\text{exp}^\pi - I_\text{tot}^\pi \end{pmatrix} \right\} \right\}, \tag{4}$$

where $\mathbf{D} = (\mathbf{PSF}^\dagger \mathbf{PSF} + \varepsilon)^{-} \mathbf{PSF}^\dagger$ with $\varepsilon = 10^{-8}$, $\mathbf{PSF}$ is the point spread function matrix and $\mathbf{D}$ the deconvolution matrix,

$$\mathbf{PSF} = \begin{bmatrix} \text{PSF}_\alpha^0 & \text{PSF}_\varphi^0 \\ \text{PSF}_\alpha^\pi & \text{PSF}_\varphi^\pi \end{bmatrix}, \tag{5}$$

The superscripts 0 and π indicate the measurements taken under either a 0° ($\text{PSF}_{\varphi,\alpha}^0$, $P^0$, $I_\text{tot}^0$, and $I_\text{exp}^0$) or 180° ($\text{PSF}_{\varphi,\alpha}^\pi$, $P^\pi$, $I_\text{tot}^\pi$, and $I_\text{exp}^\pi$) rotation.

**A quasi-Newton iterative reconstruction algorithm for XSN**

The linearized model presented in Eq. (4) is effective under weak-scattering conditions but fails to capture the nonlinear interactions in samples with strong phase shifts or absorption[21]. To overcome this limitation and improve reconstruction accuracy, we develop a quasi-Newton iterative reconstruction algorithm tailored for the XSN framework.

Rather than relying on the linear approximation, we return to the full image formation model described in Eq. (1), which accurately captures intensity formation under strong scattering. In this framework, the deconvolution result from Eq. (4) is used to construct an approximate inverse Jacobian, enabling efficient updates of the complex refractive index through quasi-Newton iterations (see also Eq. (14)).

The detailed reconstruction procedure is summarized in Algorithm 1, while a full derivation, performance benchmarking, and comparisons with alternative iterative methods is provided in the *Methods* section. Our approach shares conceptual similarities with recent iterative schemes developed for visible-light microscopy[22-24], although it differs in the formulation of the cost function and optimization strategy to accommodate the specific physics of X-ray phase imaging with partial coherence and improve convergence speed.

$$\alpha_0 = 0; \varphi_0 = 0; I_0^{0,\pi} = 0$$
for j=1:7
$$\begin{Bmatrix} \alpha_j \\ \varphi_j \end{Bmatrix} = \begin{Bmatrix} \alpha_{j-1} \\ \varphi_{j-1} \end{Bmatrix} + \max(I_{j-1}^0, I_{j-1}^\pi)^{-1} \mathcal{F}^{-1} \left\{ \mathbf{D}\mathcal{F} \left\{ \begin{pmatrix} I_{\exp}^0 - I_{j-1}^0 \\ I_{\exp}^\pi - I_{j-1}^\pi \end{pmatrix} \right\} \right\}$$
$$I_j^{0,\pi} = \int \left| \mathcal{F}^{-1} \left\{ P^{0,\pi} \mathcal{F} \left\{ \sqrt{I_\mathbf{k}} e^{i\mathbf{k}\mathbf{r}} e^{\alpha_j + i\varphi_j} \right\} \right\} \right|^2 d\mathbf{k}$$
end

Algorithm 1. Pseudocode of the XSN field reconstruction algorithm using the quasi-Newton method.

Simulating spatially incoherent light scattering is generally computationally demanding, as each incoherent illumination mode **k** in Eq. (1) must be computed individually. In diffraction-limited imaging systems where the illumination numerical aperture (NA) matches that of detection, the number of independent modes can approach the number of pixels—often exceeding one million. Such computational loads per iteration makes large-scale reconstructions impractical.

To mitigate this challenge, we leverage two key features of our optical setup. First, the illumination NA is only 22% of the detection NA (Fig. S1a), significantly reducing the number of angular modes required for accurate modeling. Second, because the pupil-plane cutoff filter is applied along a single axis (horizontal), we assume partial coherence only in the horizontal direction. This simplification reduces the number of effective illumination modes to approximately 100 for a 20 μm field of view. As a result of these optimizations, the execution time for each field reconstruction iteration is reduced to approximately 11 milliseconds.

Our algorithm proves particularly effective for samples with strong absorption. As shown in Fig. S2b, iterative reconstruction yields markedly improved tomograms, with a monotonic decrease in reconstruction error across iterations (Fig. S2c). The algorithm typically converges within seven iterations, a value used consistently for all reconstructions throughout the study. Following field reconstruction, the three-dimensional tomogram is obtained using standard filtered back projection (see *Methods*).

**Accuracy and resolution validation**

To assess the accuracy of XSN, we imaged a reference sample composed of Al, $SiO_2$, and Cu microparticles (see *Methods*). A line profile was extracted from the reconstructed tomograms to compare the experimentally measured refractive index decrement $\delta$ and absorption index $\beta$ with theoretical values (Fig. 2b). The results show good agreement, although the $\delta$ value for Cu is slightly underestimated. This discrepancy is likely due to Cu's high refractive index and strong absorption, which introduce nonlinear effects that challenge accurate reconstruction.

We also evaluated the background noise levels of $\delta$ and $\beta$ in a homogeneous region (Fig. 2a, pink square). Both values exhibit similar noise levels, consistent with the comparable strength of the optical transfer function for $\delta$ and $\beta$ in the mid-to-high spatial frequency range. However, for most materials, the signal from $\delta$ is typically an order of magnitude stronger than that from $\beta$, resulting in a significantly higher signal-to-noise ratio for the $\delta$ image. In biological samples or low-Z elements, this contrast is often even more pronounced—up to three orders of magnitude. While phase imaging provides superior sensitivity to fine structural features, the relatively weaker low-frequency response of the optical transfer function for $\delta$ may reduce precision in quantifying its absolute value.

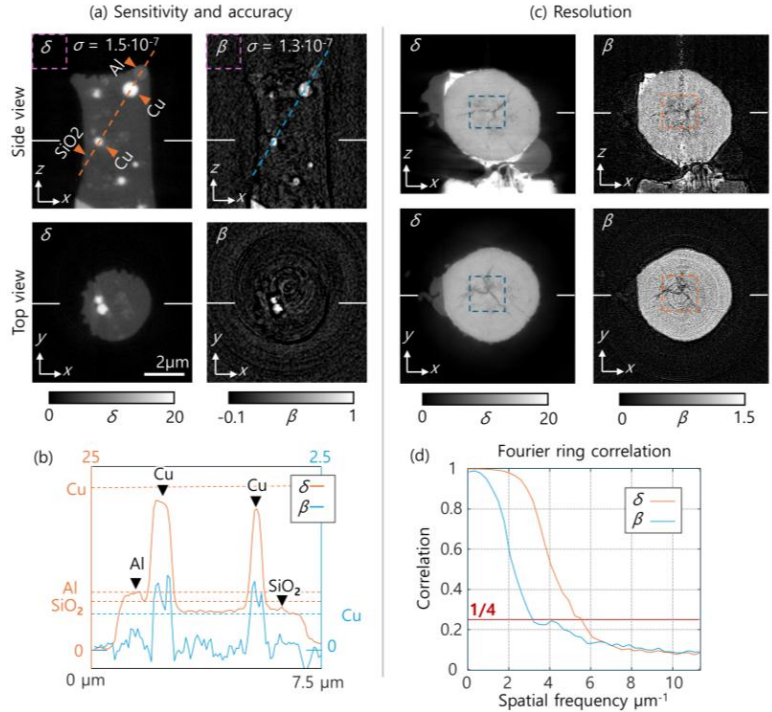

Figure 2. Analysis of resolution, sensitivity, and accuracy. (a) Reconstructed $\delta$ and $\beta$ tomograms of a mixed Al–SiO$_2$–Cu microparticle sample. (b) Line profile comparison of the measured $\delta$ and $\beta$ values with theoretical expectations. (c) XSN tomogram of a nickel–manganese–cobalt (NMC) cathode particle with visible microcracks. (d) Resolution assessment in the microcrack region (highlighted in c) using Fourier ring correlation (FRC). The phase image achieves higher resolution due to its stronger signal. $\delta$ and $\beta$ values are expressed in units of $10^{-6}$.

Combining δ and β images also enhances material specificity. In this experiment, only Cu exhibits substantial X-ray absorption and is readily identified in the β map. Meanwhile, the δ image reveals the presence and morphology of lighter elements such as Al and $SiO_2$.

Resolution is a critical metric in microscopy, particularly for applications requiring nanoscale structural characterization. To evaluate the spatial resolution of XSN, we analyzed a nickel–manganese–cobalt (NMC) battery cathode particle, which exhibits fine internal fractures resulting from electrochemical cycling (Fig. 2c). Assessing such microcracks, along with porosity, is essential for understanding and improving battery performance[25-27].

While these fractures are clearly visible in the phase δ image, the absorption β image lacks sufficient signal-to-noise ratio to resolve the smallest features. To quantify the resolution difference, we applied Fourier ring correlation (FRC), a noise-aware method that estimates spatial resolution by evaluating the correlation between two independently reconstructed tomograms across spatial frequency bands (see *Methods*; Ref.[28]).

Using a 1/4 correlation threshold, the phase image achieved a resolution of 91 nm, whereas the absorption image exhibited a lower resolution of 158 nm (Fig. 2d). The resolution of the absorption channel is primarily limited by photon shot noise, while the phase resolution is constrained by the scintillator's point spread function, which has a full width at half maximum (FWHM) of 72 nm (see *Methods*).

These results highlight the intrinsic advantage of phase imaging for resolving fine structural features. An additional benefit is that phase contrast can be obtained at X-ray energies away from absorption edges, thereby minimizing radiation-induced damage to battery materials or biological specimens[29,30].

**Hyperspectral imaging of battery cathodes particle**

In the previous section, we demonstrated that phase imaging is highly effective for visualizing the structural features of NMC (nickel–manganese–cobalt) battery particles. However, for practical applications such as performance evaluation and degradation analysis, chemical composition is often of greater relevance. In X-ray imaging, absorption is directly influenced by atomic composition and electronic states[31]. Notably, each element exhibits a characteristic K-edge—the energy required to eject a core (K-shell) electron—which provides valuable information about its identity and oxidation state. Analysis of the K-edge structure, including the pre- and post-edge regions, can further reveal chemical bonding and valence states[32].

NMC cathode particles are primarily composed of Ni, Co, and Mn. The K-edges of these elements are located at 8333 eV (Ni), 7709 eV (Co), and 6539 eV (Mn), respectively[33]. To explore the hyperspectral capabilities of XSN, we performed energy-resolved imaging across the Ni K-edge (Fig. 3). The undulator gap, monochromator, and zone plate were automatically adjusted using motorized stages to accommodate each X-ray energy step, while the remaining optical components—including the condenser—remained fixed throughout the scan.

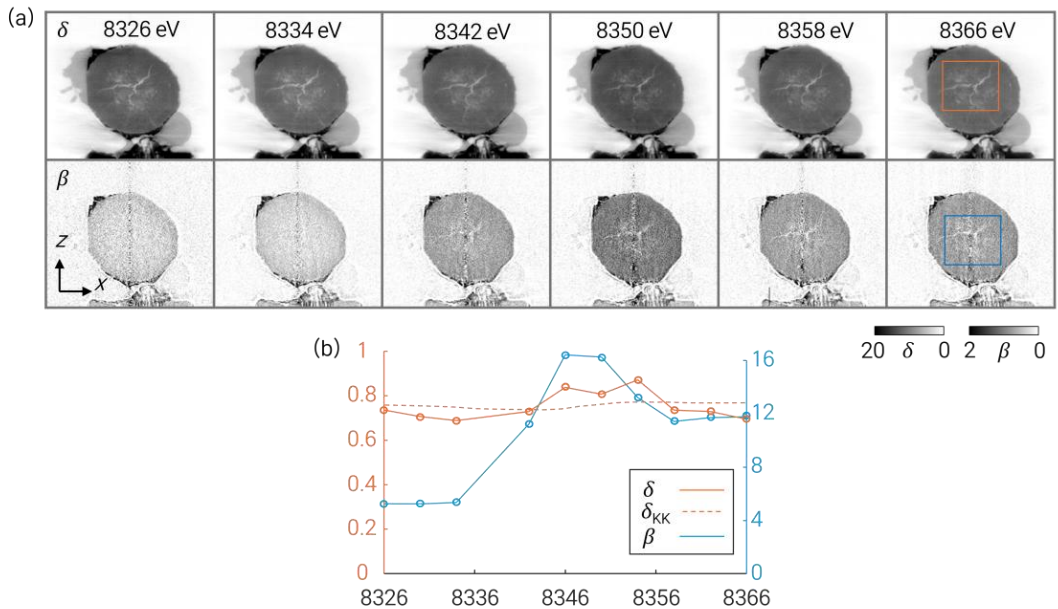

Figure 3. (a) Hyperspectral phase and absorption images of the gradient NMC sample acquired across the Ni K-edge. (b) Plots of average $\delta$ and $\beta$ values in selected regions (marked in a), compared to the expected $\delta$ profile derived from Kramers–Kronig relations. $\delta$ and $\beta$ index values are reported in units of $10^{-6}$.

The Ni K-edge is marked by a sharp increase in X-ray absorption, which is clearly visible in our experimental data (Fig. 3a). As is characteristic of nickel oxides[34], the absorption peaks just above the edge—within the near-edge region—before gradually decreasing. This near-edge structure arises from fine features in the local atomic and electronic environment surrounding the absorbing atom, and is widely utilized for chemical state analysis in battery imaging[27,35-38].

In theory, the corresponding phase signal $\delta$ is expected to exhibit a modest decrease in response to the increase in absorption $\beta$ at the K-edge[15,39], with the relative change being proportional to the scale of the variation in the absorption image. However, in our case, the mean $\delta$ value is nearly an order of magnitude larger than the peak $\beta$ value. As a result, the phase remains effectively constant across the scanned energy range, as confirmed by the experimental data in Fig. 3b.

The consistency of the measured phase and absorption values can be evaluated using Kramers–Kronig relations applied to the measured $\beta$ spectrum (Fig. 3b)[40]. The theoretical prediction yields an almost flat phase response across the scanned energy window, closely matching our observations. Minor deviations from the theoretical curve are likely due to imprecisions in the low spatial frequency components of the phase reconstruction, uncertainties in the cutoff filter alignment, and residual phase–intensity coupling effects introduced by stray light, which can elevate the black level.

We next quantified the elemental composition of Ni, Co, and Mg within the cathode particles. To avoid artifacts arising from near-edge structures, we used tomograms acquired further before (8326 eV) and after (8366 eV) the Ni K-edge to compute chemical densities based on Eqs. (17) and (18) (see *Methods*).

In the gradient NMC particle, we observed a gradual decrease in Co and Mg concentrations from the outer edge toward the center (Fig. 4b), consistent with the intended compositional gradient. In contrast, nickel–

cobalt–aluminum (NCA) particles exhibited uniform elemental distributions, reflecting their nominal composition with ~11% Co.

The chemical density profile of the gradient NMC particle was further validated against transmission electron microscopy with energy-dispersive X-ray spectroscopy (TEM-EDS), performed on a particle prepared under identical conditions (Fig. S3). The Mn+Co to Ni ratio was ~0.5 at the particle's periphery and decreased toward the center, in good agreement with the XSN-based measurements.

Some artifacts observed at the sample edges in the density map are attributed to platinum (Pt) deposited during sample mounting for focused ion beam preparation. Additionally, we found that the charged NCA particle exhibited more internal fractures and a lower overall density compared to the pristine (discharged) counterpart (Fig. 4a–b), in line with known degradation effects induced by electrochemical cycling[25].

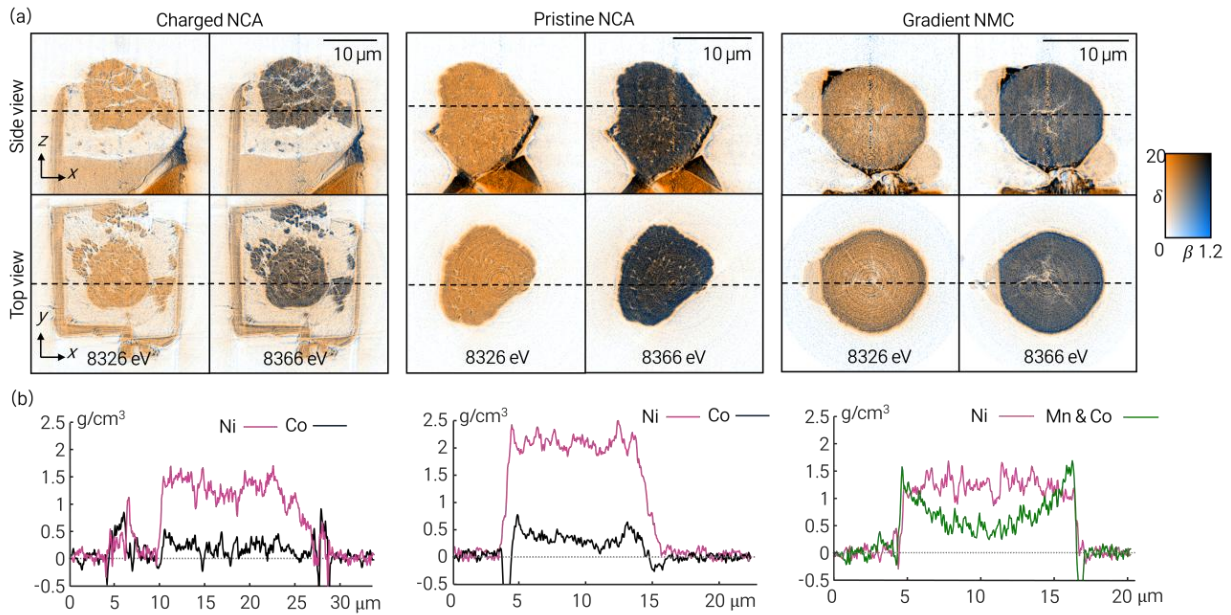

Figure 4. a) Tomograms of different NCA and NMC at a wavelength before and after the k-edge $\delta$ is displayed in the orange channel and $\beta$ in the blue one (refer to the colormap for colormixing). b) Derived chemical composition analysis of various NCA and NMC samples along the dotted lines in the tomograms in (a). $\delta$ and $\beta$ index values are reported in units of $10^{-6}$

**Conclusion**

We developed XSN (quantitative X-ray schlieren nanotomography) as a fast, experimentally accessible approach for quantitative three-dimensional imaging of both refractive index and absorption using hard X-rays. Building on the optical design of our earlier Kramers–Kronig nanotomography[16], XSN introduces two key advances: a schlieren-inspired, partially coherent illumination scheme with an asymmetric pupil-plane filter for directional phase encoding, and a quasi-Newton iterative inversion algorithm for accurate reconstruction of the complex refractive index. Together, these innovations enable robust, high-resolution phase imaging under realistic experimental conditions (including partial coherence and strong absorption) while preserving a simple full-field microscope setup.

A distinguishing feature of XSN is its ability to simultaneously retrieve phase and absorption from each single projection, without requiring multiple exposures, fringe pattern encoding, interferometric stability, or weak-object approximations. This single-shot quantitative phase retrieval remains effective even for optically thick or strongly absorbing specimens that would defy conventional phase-contrast approaches. By explicitly incorporating partial coherence into the forward model and employing a robust iterative solver, XSN achieves accurate, stable reconstructions across a broad range of sample types and imaging conditions.

Furthermore, XSN is inherently compatible with multi-energy (spectral) imaging. It employs an achromatic phase-contrast filter (a flat silicon wafer edge) so that phase contrast is maintained over a range of X-ray energies with minimal realignment. This enables streamlined acquisition of hyperspectral four-dimensional datasets (combining 3D spatial imaging with energy resolution), providing simultaneous structural and chemical information.

We validated XSN's quantitative accuracy and spatial resolution using reference microspheres of Al, Cu, and $SiO_2$. The retrieved refractive index values closely matched theoretical predictions, and the method achieved a phase spatial resolution of approximately 91 nm—substantially finer than that of absorption-based imaging. This performance underscores the enhanced sensitivity of phase contrast for resolving nanoscale features.

As a demonstration on complex real-world specimens, we applied XSN to lithium-ion battery cathodes. In nickel–manganese–cobalt (NMC) particles, XSN revealed internal microcracks and, by scanning across the Ni K-edge, mapped the three-dimensional distribution of chemical composition. This combined structural and compositional imaging is particularly valuable for energy storage materialsfile[41], where internal degradation and elemental gradients critically influence performance and lifespan.

Beyond the battery example, XSN's unique combination of high sensitivity, spectral flexibility, and minimal hardware complexity makes it broadly applicable across diverse fields. The approach can be readily adapted to biological imaging[42], advanced materials characterization, and semiconductor inspection[43], providing a powerful means to nondestructively probe internal structures at the nanoscale. Equally important, XSN requires no specialized optics or elaborate alignment, enabling straightforward implementation on standard synchrotron beamlines and seamless integration into existing microscopy facilities.

Finally, the quasi-Newton iterative reconstruction framework developed here has broad relevance beyond X-ray imaging. It can be generalized to other modalities—including visible light, electron, and terahertz imaging—where partial coherence and multiple scattering similarly impede quantitative reconstruction.

In summary, quantitative X-ray schlieren nanotomography provides a powerful and versatile platform for high-resolution X-ray imaging, enabling simultaneous three-dimensional mapping of structure and composition across a range of photon energies. By uniting the simplicity of single-shot full-field acquisition with the rigor of advanced computational phase retrieval, XSN bridges the gap between rapid data collection and precise nanoscale quantification. This integrated capability opens new avenues for non-destructive, high-throughput investigations in both the physical and life sciences, with far-reaching applications in materials science, energy technology, and biomedical research.

**Methods**

**Optical setup**

Imaging is performed at the X-ray nanoimaging (7C, XNI) beamline of the Pohang Light Source II (PLS-II). An undulator source with a liquid-nitrogen-cooled silicon (111) double-crystal monochromator (DCM) (Vactron Co., Ltd., Republic of Korea) is used to obtain temporal coherence ($\Delta E / E \cong 10^{-4}$). A flat mirror is introduced to remove the harmonic frequencies.

A compound refractive lens (CRL) focuses the generated X-ray illumination onto the sample. The CRL is consists of three beryllium lenses with a radius of curvature of 0.05 mm (RXOPTICS GmbH, Germany). A diffuser (Advantec filter paper No.101) is placed near the pupil plane of the illumination to achieve the Köhler illumination condition. Such a configuration also uniformizes the illumination without increasing the illumination NA[18], which is good for sensitivity at low spatial frequencies. In our case, the rotating diffuser is placed directly after the CRL lenses. A pinhole is placed directly in front of the sample to limit the illumination size on the sample.

The imaging part of the setup follows an off-axis configuration making it easily reproducible at any synchrotron beamline. A detailed schematic of the setup with the distances between the elements are shown in Fig. S1b. The sample diffraction is collected by a zone plate (Applied Nanotools, 300um diameter, 30nm outermost, Au, 800nm Thickness). A silicon cutoff filter is positioned at the pupil plane of the zone plate. The filter is a 0.5-mm-thick silicon wafer with the flat top used as the cutoff edge. Its axial position is determined using the Foucault knife edge test[44]: finding the position where moving the edge into the beam uniformly reduces its intensity. The beam intensity is measured using a with 20-um-thick Ce:GAGG scintillator and magnified using an objective and tube lens system (Zeiss LD Plan-Neofluar 20x/0.4 Corr M27) before being detected by the camera (Teledyne 01-KINETIX-M-C, 2x2 binning, 2 s exposure). Images are acquired on the fly while the sample is rotated at a rate of 0.2 degrees/second. A total of 900 images are acquired for a tomogram, which takes about 35 minutes. Hyperspectral imaging was conducted across the Ni K-edge (8333 eV) by increasing the energy in a 40 eV range with 4 eV steps, the zone plate was moved backward toward the sample with a motorized stage at a rate of 4.8um/eV to maintain focus.

**Sample preparation**

The microparticle mixture sample shown in Fig. 2 is prepared by mixing three types of microparticles in a volume ratio of 1:1:1. The three types of microparticles are silica (200 nm, US1151M, US Research Nanomaterials, Inc., USA), aluminum (800 nm, US1042, US Research Nanomaterials, Inc., USA), and copper (300 nm, US5002, US Research Nanomaterials, Inc., USA). Microparticles are bound with cyanoacrylate adhesive (Loctite 401, Henkel) and applied to a tungsten tip (T20- 7B, Modusystems, Inc., Republic of Korea). After curing, the sample was cut into shape using a focused ion beam system.

The full cell single-particle NCA (pristine and charged), composed of $LiNi_{0.88}Co_{0.1}Al_{0.01}O_2$ (NCA) and graphite, was charged to 4.2 V at 0.2 C-rate during the third cycle to obtain the charged NCA. The cells were then disassembled, and the residual electrolyte was removed using dimethyl carbonate. Gradient $LiNi_xCo_yMn_zO_2$ (NCM) with a nominal composition of Ni:Co:Mn = 60:20:20 was synthesized using a conventional co-precipitation methods. The outer region consists of approximately 90% Ni, with Co and Mn each around 5%, while the core contains about 50% Ni, 30% Co, and 20% Mn (Fig. S3). Single-particle NCA (pristine and charged) and gradient NCM for QSXCT imaging were prepared using a Focused Ion Beam Helios 600 at the National Institute for Nanomaterials Technology (NINT), POSTECH.

**Angular power spectral density measurement**

The calibration of the angular power spectral density $I_k$ is an important prerequisite for the calculation of point spread function in Eq. (10). However, it is difficult to measure directly in practice due to the bulkiness of the detector, the risk of beam damage, and the large pixel size. Therefore, we calibrated $I_k$ indirectly by measuring the mean background intensity while scanning the cutoff filter position. The angular power spectral density is then computed from the derivative of the mean background intensity by the cutoff position (Fig. S1a(i)),

$$I_{\mathbf{k}=(k_x,0)} = \frac{\langle I(x_{cut})\rangle_{bg}}{\partial x_{cut}}, \quad (6)$$

where $\langle I(x_{cut})\rangle_{bg}$ is the spatial average of the background intensity for the cutoff filter position $x_{cut}$, $k_x = 2\pi x_{cut}/(\lambda f_{ZP})$ is the horizontal component of the angular wavevector at the sample plane, $\lambda$ is the wavelength, and $f_{ZP}$ is the distance from the zone plate to the cutoff filter. Once the pupil plane intensity is known, the optical transfer function can be computed [Eq. (10)] and the field reconstructed (Algorithm 1, Fig. S1a(ii)).

**Linearization of the image formation model**

Equation (1) can be linearized around $\alpha(\mathbf{r}) + i\varphi(\mathbf{r})$ with a Taylor expansion for a perturbation $\alpha'(\mathbf{r}) + i\varphi'(\mathbf{r})$ with $\alpha'(\mathbf{r}), \varphi'(\mathbf{r}) \ll 1$,

$$\begin{aligned}I(\mathbf{r}) &= \int \left|\mathcal{F}^{-1}\left\{P(\mathbf{u})\mathcal{F}\left\{\sqrt{I_{\mathbf{k}}}e^{i\mathbf{kr}}e^{\alpha(\mathbf{r})+i\varphi(\mathbf{r})+\alpha'(\mathbf{r})+i\varphi'(\mathbf{r})}\right\}\right\}\right|^2 d\mathbf{k} \\ &\underset{\text{taylor}}{\cong} \int \left|\mathcal{F}^{-1}\left\{P(\mathbf{u})\mathcal{F}\left\{\sqrt{I_{\mathbf{k}}}e^{i\mathbf{kr}}e^{\alpha(\mathbf{r})+i\varphi(\mathbf{r})}\left(1+\alpha'(\mathbf{r})+i\varphi'(\mathbf{r})\right)\right\}\right\}\right|^2 d\mathbf{k}\end{aligned} \quad (7)$$

We first study the case $\alpha(\mathbf{r}) + i\varphi(\mathbf{r}) = 0$. In this case equation (7) reduces to

$$I(\mathbf{r}) \cong I_{tot} + \int_{\mathbf{k}} 4I_{\mathbf{k}} \operatorname{Re}\left[e^{-i\mathbf{kr}}P^*(\mathbf{k})\mathcal{F}^{-1}\left\{P(\mathbf{u})\mathcal{F}\left\{e^{i\mathbf{kr}}\left(\alpha'(\mathbf{r})+i\varphi'(\mathbf{r})\right)\right\}\right\}\right]d\mathbf{k} \quad (8)$$

Where only the first order of $\alpha'(\mathbf{r}), \varphi'(\mathbf{r})$ were kept and $I_{tot} = \sum_{\mathbf{k}} I_{\mathbf{k}}$ is the intensity in the absence of sample. Using the Fourier shift theorem, the equation further reduces to

$$I(\mathbf{r}) \cong I_{tot} + \int_{\mathbf{k}} 4\operatorname{Re}\left[\mathcal{F}^{-1}\left\{I_{\mathbf{k}}P^{*}(\mathbf{k})P(\mathbf{u}-\mathbf{k})\mathcal{F}\{\alpha'(\mathbf{r})+i\varphi'(\mathbf{r})\}\right\}\right]d\mathbf{k}. \tag{9}$$

This equation can be interpreted as a combination of convolutions

$$\begin{aligned}
I &= I_{tot} + \mathrm{PSF}_{\alpha} * \alpha' + \mathrm{PSF}_{\varphi} * \varphi', \\
\mathrm{PSF}_{\alpha} &= 4\operatorname{Re}\left[\mathcal{F}^{-1}\{I_{\mathbf{k}}P\}^{*}\mathcal{F}^{-1}\{P\}\right], \\
\mathrm{PSF}_{\varphi} &= 4\operatorname{Im}\left[\mathcal{F}^{-1}\{I_{\mathbf{k}}P\}^{*}\mathcal{F}^{-1}\{P\}\right],
\end{aligned} \tag{10}$$

where the parameter $\mathbf{r}$ of $\alpha$ and $\varphi$ is omitted for clarity. This convolution formulation of image formation is widely used in visible light microscopy to quantitatively reconstruct differential phase contrast measurements[19,45]. Furthermore, the approximation $\alpha'(\mathbf{r}), \varphi'(\mathbf{r}) \ll 1$ used in the derivation has been found to also be valid for larger phase values of $\varphi'$ which are slowly varying[21]. If we carry two intensity measurement $I_{\exp}^{0}$ and $I_{\exp}^{\pi}$ with corresponding PSFs $\mathrm{PSF}_{\alpha,\varphi}^{0}$ and $\mathrm{PSF}_{\alpha,\varphi}^{\pi}$, the intensity can then be deconvolution with a Tikhonov regularized deconvolution [20,46]

$$\begin{pmatrix} \alpha'(\mathbf{r}) \\ \varphi'(\mathbf{r}) \end{pmatrix} = \mathcal{F}^{-1}\left\{\mathbf{D}\mathcal{F}\left\{\begin{pmatrix} I_{\exp}^{0} \\ I_{\exp}^{\pi} \end{pmatrix}\right\}\right\}$$
$$\mathbf{PSF} = \begin{pmatrix} \mathrm{PSF}_{\alpha}^{0} & \mathrm{PSF}_{\varphi}^{0} \\ \mathrm{PSF}_{\alpha}^{\pi} & \mathrm{PSF}_{\varphi}^{\pi} \end{pmatrix}; \mathbf{D} = (\mathbf{PSF}^{?}\mathbf{PSF} + \varepsilon)^{-}\mathbf{PSF}^{?} \tag{11}$$

Since we are interested in iterative reconstruction the case $\alpha(\mathbf{r}) + i\varphi(\mathbf{r}) \neq 0$ is also of interest. Solving the linearized equation around a given point $\alpha(\mathbf{r}) + i\varphi(\mathbf{r})$ corresponds to a Gauss-Newton iteration. In this case equation (7) reduces to

$$I(\mathbf{r}) \cong I_{\alpha,\varphi} + \int_{\mathbf{k}} 4\operatorname{Re}\left[I_{\mathbf{k}}\mathcal{F}^{-1}\left\{P(\mathbf{u})\mathcal{F}\left\{e^{i\mathbf{k}\mathbf{r}}e^{\alpha(\mathbf{r})+i\varphi(\mathbf{r})}\right\}\right\}^{*}\mathcal{F}^{-1}\left\{P(\mathbf{u})\mathcal{F}\left\{e^{i\mathbf{k}\mathbf{r}}e^{\alpha(\mathbf{r})+i\varphi(\mathbf{r})}(\alpha'(\mathbf{r})+i\varphi'(\mathbf{r}))\right\}\right\}\right]d\mathbf{k} \tag{12}$$

with $I_{\alpha,\varphi} = \int_{\mathbf{k}} \left| \mathcal{F}^{-1}\left\{ P(\mathbf{u}) \mathcal{F}\left\{ \sqrt{I_\mathbf{k}} e^{i\mathbf{k}\mathbf{r}} e^{\alpha(\mathbf{r})+i\varphi(\mathbf{r})} \right\} \right\} \right|^2 d\mathbf{k}$ the intensity image formed for a sample of complex phase $\alpha(\mathbf{r}) + i\varphi(\mathbf{r})$. This equation is much harder to solve. One approximation that can facilitate solving this equation is that $e^{\alpha(\mathbf{r})+i\varphi(\mathbf{r})}$ is mainly composed of low frequency components a typical feature of transmitted phase image. In that case $e^{\alpha(\mathbf{r})+i\varphi(\mathbf{r})}$ can be considered unaffected by the cutoff filtering and can be commuted with the filtering operation:

$$I(\mathbf{r}) \cong I_{\alpha,\varphi} + \int_{\mathbf{k}} 4\operatorname{Re}\left[ (I_\mathbf{k} P(\mathbf{k})) e^{-i\mathbf{k}\mathbf{r}} \mathcal{F}^{-1}\left\{ P(\mathbf{u}) \mathcal{F}\left\{ e^{i\mathbf{k}\mathbf{r}} e^{2\alpha(\mathbf{r})} (\alpha'(\mathbf{r}) + i\varphi'(\mathbf{r})) \right\} \right\} \right] d\mathbf{k}, \tag{13}$$

which can be reformulated as a convolution

$$I = I_{\alpha,\varphi} + \mathrm{PSF}_\alpha * (e^{2\alpha} \alpha') + \mathrm{PSF}_\varphi * (e^{2\alpha} \varphi'),$$
$$\mathrm{PSF}_\alpha = 4\operatorname{Re}\left[ \mathcal{F}^{-1}\{I_\mathbf{k} P\}^* \mathcal{F}^{-1}\{P\} \right],$$
$$\mathrm{PSF}_\varphi = 4\operatorname{Im}\left[ \mathcal{F}^{-1}\{I_\mathbf{k} P\}^* \mathcal{F}^{-1}\{P\} \right],$$

and inversed accordingly

$$\begin{pmatrix} \alpha' \\ \varphi' \end{pmatrix} = e^{-2\alpha} \mathcal{F}^{-1}\left\{ \mathbf{D} \mathcal{F}\left\{ \begin{pmatrix} I_{\exp}^0 - I_{\alpha,\varphi}^0 \\ I_{\exp}^\pi - I_{\alpha,\varphi}^\pi \end{pmatrix} \right\} \right\},$$

where the parameter $\mathbf{r}$ of $\alpha$ and $\varphi$ is omitted for clarity. $e^{-2\alpha}$ can be understood as the transmitted intensity in the absence of filter. In practice we found that replacing $e^{-2\alpha}$ by $\max(I_{\alpha,\varphi-}, I_{\alpha,\varphi+})^{-1}$ the maximum transmitted intensity in the presence of a filter made the algorithm more stable.

$$\begin{pmatrix} \alpha' \\ \varphi' \end{pmatrix} = \max(I_{\alpha,\varphi}^0, I_{\alpha,\varphi}^\pi)^{-1} \mathcal{F}^{-1}\left\{ \mathbf{D} \mathcal{F}\left\{ \begin{pmatrix} I_{\exp}^0 - I_{\alpha,\varphi}^0 \\ I_{\exp}^\pi - I_{\alpha,\varphi}^\pi \end{pmatrix} \right\} \right\}. \tag{14}$$

This linear image formation model can be used as an approximation of the inverse Jacobian in the Gauss-Newton method which will be discussed in the next section.

**Iterative reconstruction algorithms**

The fields are reconstructed using the Gauss-Newton algorithm. At each step the linear equation (12) can be solved by explicitly formulating it as a matrix multiplication $\begin{pmatrix} I_{exp}^0 - I_{\alpha,\varphi}^0 \\ I_{exp}^\pi - I_{\alpha,\varphi}^\pi \end{pmatrix} = \mathbf{J} \begin{pmatrix} \alpha' \\ \varphi' \end{pmatrix}$ with the matrix J corresponding to the Jacobian matrix of equation (1). After inversion the iterative Gauss-Newton update is

$$\begin{pmatrix} \alpha' \\ \varphi' \end{pmatrix} = (\mathbf{J}^?\mathbf{J} + \varepsilon)^{-} \mathbf{J} \begin{pmatrix} I_{exp}^0 - I_{\alpha,\varphi}^0 \\ I_{exp}^\pi - I_{\alpha,\varphi}^\pi \end{pmatrix}$$
$$\begin{pmatrix} \alpha \\ \varphi \end{pmatrix} \to \begin{pmatrix} \alpha + \alpha' \\ \varphi + \varphi' \end{pmatrix}. \qquad (15)$$

The matrix J is the sum of large dense matrices and has to be computed every iteration resulting in a long execution time (see "Exact Newton" in Fig. S2a).

One way to speed up iterations[24] is to fix the Jacobian at the first iteration—here corresponding to solving Eq. (11)—this can accelerate the execution time at the cost of slower convergence (see "Quasi-Newton" in Fig. S2a). Finally, when using the improved Jacobian approximation normalized with the maximum intensity given in Eq. (14), the execution time remains fast while also achieving fast convergence (see "Quasi-Newton normalized" in Fig. S2a). For this case the full iterative algorithm is presented in the Results section (see Algorithm 1) and is used throughout the paper.

The linear non-iterative algorithm [Eq. (11)] is also benchmarked in Fig. S2a. The linear algorithm is the fastest because it does not require any iteration (Fig. S2a). However, it becomes inaccurate for strongly absorbing objects, resulting in an underestimation of both the phase and absorption tomograms (Fig. S2b). In all algorithms, the phase and absorption spatial spectra were trimmed to the $\mathcal{F}\{\text{PSF}_\alpha\} < 0.8$ at every iteration to avoid high frequency noise

**Scintillator modulation transfer function deconvolution**

To obtain a precise measurement of the X-ray field intensity the camera image must be deconvolved considering scintillator thickness and visible light optics between the scintillator and camera. We determine the pupil function $P_{sc}$ at a given depth $z_{sc}$ inside the scintillator of refractive index $n_{sc}$ and thickness $t_{sc}$ as : $P_{sc}(k_x, k_y, z_{sc}) = \exp\left(i2\pi z_{sc}\sqrt{k_x^2 + k_y^2 + n_{sc}/\lambda^2}\right)$. The optical transfer function $\text{OTF}_{sc}$ is estimated at a given depth as the autocorrelation of the pupil function, $\text{OTF}_{sc}(k_x, k_y, z) = \mathcal{F}^{-1}\left\{|\mathcal{F}\{P_{sc}(k_x, k_y, z)\}|^2\right\}$. Finally, the optical transfer function is averaged inside the scintillator, $\text{OTF}_{sc2D}(k_x, k_y) = \int_{-t_{sc}/2}^{+t_{sc}/2} \text{OTF}_{sc}(k_x, k_y, z)dz$ and used to deconvolve the raw data using Wiener deconvolution. The point spread function calculated for our scintillator parameters has a full width at half maximum of 72 nm.

**Resolution assessment with Fourier ring correlation**

Resolution is evaluated using Fourier ring correlation. Reconstructed sinograms were divided into even and odd projections. The field sinograms $\text{sino}_{1,2}$ are then filtered in the angular direction to remove ring artifacts $\mathcal{F}_{\theta,r}^{-1}\{\mathcal{F}_{\theta,r}\{\text{sino}_{1,2}\}e^{k_r^2/\sigma_r^2}\}$. Here $\theta, r$ are the axial and radial coordinate, $\mathcal{F}_{\theta,r}\{\cdot\}$ the Fourier transform in cylindrical coordinates, $k_r$ the radial angular vector and $\sigma_r = 1$ μm$^{-1}$. Each sinogram $\text{sino}_{1,2}$ is used to reconstruct two tomograms $\delta_1, \beta_1$ and $\delta_2, \beta_2$ with Blackman window to avoid edge artifacts. The two tomograms are then compared using the usual Fourier ring correlation formula

$$FRC(\delta_1, \delta_2, \kappa) = \frac{\langle \mathcal{F}\{\delta_1\}\mathcal{F}\{\delta_2\}^*\rangle_{|\kappa - k_{xy}|<\Delta\kappa/2}}{\sqrt{\langle \mathcal{F}\{\delta_1\}\mathcal{F}\{\delta_1\}^*\rangle_{|\kappa - k_{xy}|<\Delta\kappa/2} \langle \mathcal{F}\{\delta_2\}\mathcal{F}\{\delta_2\}^*\rangle_{|\kappa - k_{xy}|<\Delta\kappa/2}}}, \quad (16)$$

where $\langle \cdot \rangle$ the average, $k_{xy}$ is the norm of the lateral wavevectors components and $\kappa \in [0, 3 \text{ μm}^{-1}], d\kappa = 0.3$ μm$^{-1}$ the ring radius and thickness.

**Automatic image alignment**

Images taken at a given sample rotation (θ) and their flipped counterpart measured at θ+180 are used to reconstruct the complex transmittance of the object (Algorithm 1). For the algorithm to work properly the two images need to be properly aligned. The optimal position is found by maximizing high frequency content after subtraction of an image and its flipped counterpart thereby maximizing the fine details in the phase image dominant in most X-ray samples. The optimization is carried out on the cost function $\left\| \mathcal{F}_{\theta,r}\{I_{exp+} - I_{exp-}\}(k_r < 0.3)k_r^2 \right\|_2^2$ using the FISTA algorithm[47] and automatic differentiation. This method can however reduce high frequency detail in the absorption image. To solve this problem, we then do a golden-section search to find the global offset between flipped images which leads to a maximization of the flatness of the Fourier ring correlation between the real and imaginary part of the reconstructed tomogram. When computing the Fourier ring correlation $FRC$, formula (16) is used on $\delta, \beta$ instead of $\delta_1, \delta_2$ and is computed for the range $\kappa \in [0, 1\ \mu m^{-1}]$. Flatness is evaluated with the cost function $\left\langle \left| FRC(\delta, \beta, \kappa) / \left\langle FRC(\delta, \beta, \kappa) \right\rangle_\kappa - 1 \right| \right\rangle_\kappa$ where $\langle \cdot \rangle_\kappa$ denotes the averaging for ring radiuses.

Image registration is one of the main challenges in reconstruction. The challenge of aligning images at 0 and 180 degrees arises from the fact that they contain complementary but non-overlapping information. Indeed, each image contains information corresponding to either the right or the left of the Fourier plane of the transmitted light field. Since both half parts contain independent information they can't be directly compared for alignment.

Improvements in the scanning scheme or setup can further facilitate image alignment and improve the accuracy of the reconstructed refractive index. For example, the use of laser interferometry could allow the sample to be precisely located in the lateral direction[48]. Acquisition of images without the cutoff filter could also facilitate alignment.

**Tomographic reconstruction**

From the transmitted fields, the tomogram is reconstructed using standard filtered back projection using the iradon function in MATLAB. The reconstructed tomogram is normalized by the constant $\lambda/(2\pi p)$ to obtain the complex refractive index $n = 1 - \delta + i\beta$ of the sample, where $p$ is the image pixel size and $\lambda$ is the wavelength.

**Atomic density measurement**

In order to avoid the near-edge structures when deriving the chemical density of Ni, Co, and Mg, we use the tomogram before (8326 eV) and after the edge (8366 eV). Because the measurements are caried far from the Co and Mg K-edges, the absorption from Co and Mg is nearly constant. Taking advantage of this, the atomic density of Ni $\sigma_{Ni}$ can be obtained by dividing the difference in the absorption tomograms $\beta$ before and after the edge, based on the reference absorption $\beta_{Ni,Ref}$ and density $\sigma_{Ni,Ref}$ data given in the literature [49].

$$\begin{aligned}
\sigma_{Ni} &= \sigma_{Ni,Ref} \frac{\beta(8326\text{eV}) - \beta(8366\text{eV})}{\beta_{Ni,Ref}(8326\text{eV}) - \beta_{Ni,Ref}(8366\text{eV})} \\
\sigma_{Ni,Ref} &= 8.9 \text{g/cm}^3 \\
\beta_{Ni,Ref}(8326\text{eV}) &= 4.5 \times 10^{-7} \\
\beta_{Ni,Ref}(8366\text{eV}) &= 3.4 \times 10^{-6}
\end{aligned} \quad (17)$$

Subsequently, for the NMC sample the combined density of Co and Mg $\sigma_{MgCo}$ was obtained by multiplying the pre-edge absorption tomogram by the average density to absorption ratio obtained from Ref.[49]. For the NCA sample only Co density was taken into account:

$$\sigma_{MgCo} = \left( \beta(8326\text{eV}) - \sigma_{Ni} \frac{\beta_{Ni,Ref}(8326\text{eV})}{\sigma_{Ni,Ref}} \right) \frac{1}{2} \left( \frac{\sigma_{Mn,Ref}}{\beta_{Mn,Ref}(8326\text{eV})} + \frac{\sigma_{Co,Ref}}{\beta_{Co,Ref}(8326\text{eV})} \right)$$

$$\sigma_{Co} = \left( \beta(8326\text{eV}) - \sigma_{Ni} \frac{\beta_{Ni,Ref}(8326\text{eV})}{\sigma_{Ni,Ref}} \right) \frac{\sigma_{Co,Ref}}{\beta_{Co,Ref}(8326\text{eV})} \quad (18)$$

$$\frac{\sigma_{Mn,Ref}}{\beta_{Mn,Ref}(8326\text{eV})} = 3.4 \times 10^6 \text{ g/cm}^3$$

$$\frac{\sigma_{Co,Ref}}{\beta_{Co,Ref}(8326\text{eV})} = 2.9 \times 10^6 \text{ g/cm}^3$$

Note that these equations approximate the combined Mn, Co density using the fact that density to absorption ratios of Mn and Co are similar. A more precise and independent quantification of Mn and Co would require additional imaging at the Mn or Co K-edge. For NCA particles, Al composition is not considered as Al atomic fraction and absorption are both one order of magnitude lower than that of Co, similarly Li and O absorption are not considered, being at least three orders of magnitude lower than Co[49].

Acknowledgement

This work was supported by National Research Foundation of Korea grant funded by the Korea government (MSIT) (RS-2024-00442348, 2022M3H4A1A02074314, 2021R1C1C2009220).

Competing interests

All other authors declare no competing interests.